\documentclass[twocolumn,english,aps,floatfix,superscriptaddress,amsfonts,amssymb,superscriptaddress]{revtex4}
\usepackage{color}

\usepackage{graphicx}
\usepackage{amsmath}
\usepackage{latexsym}\bfseries
\usepackage{epstopdf}
\usepackage{amsmath}
\usepackage{amssymb,amsmath}
\usepackage{multirow}
\usepackage{mathtools}
\newcommand{\beq}{\begin{equation}}
\newcommand{\eeq}{\end{equation}}
\newcommand{\tn}{\textbf}

\date{\today{}}

\def\ket#1{|#1\rangle}

\usepackage{tikz}
\usepackage{xcolor}
\usetikzlibrary{patterns}

\begin{document}

\title{Multipartite entanglement and quantum Fisher information in conformal field theories}

\author{M.~A.~Rajabpour}
\affiliation{  Instituto de F\'isica, Universidade Federal Fluminense, Av. Gal. Milton Tavares de Souza s/n, Gragoat\'a, 24210-346, Niter\'oi, RJ, Brazil}
\begin{abstract}
Bipartite entanglement entropy of a segment with the length $l$ in $1+1$ dimensional conformal field theories (CFT) follows the formula $S=\frac{c}{3}\ln l+\gamma$, where $c$
is the central charge of the CFT and $\gamma$ is a cut-off dependent constant which diverges in the absence
of an ultraviolet cutoff. According to this formula, systems with larger central charges have {\it{more}} bipartite entanglement entropy. Using quantum Fisher 
information (QFI), we argue that
systems with bigger central charges not only have larger bipartite entanglement entropy but also have more multipartite entanglement content. In particular, we argue that
since systems with smaller {\it{smallest scaling dimension}} have bigger QFI, the multipartite entanglement content of a CFT is dependent on the value of 
the smallest scaling dimension present in the spectrum of the system. We show that our argument seems to be consistent with some of the existing results 
regarding the von Neumann entropy, negativity, and localizable entanglement in $1+1$ dimensions. 
Furthermore, we also argue that the QFI decays under renormalization group (RG) flow between two unitary CFTs. 
Finally, we also comment on the non-conformal but scale invariant systems.

\end{abstract}
\maketitle
\section{Introduction}

Understanding quantum field theories (QFT) based on their entanglement content has been one of the most active lines of research in
the last few decades. Entanglement entropy is one of the most studied bipartite entanglement measures, and it has been investigated in great detail in many quantum field theories.
Strictly speaking, entanglement entropy in quantum field theories is a cutoff dependent quantity\cite{Sorkin,CW,Holzhy}. However,
the interesting observation is that the divergence of this quantity in some cases is related to the universal structure of the quantum field theory. For example, 
in $1+1$ dimensional conformal field theories (CFT) the entanglement entropy of a segment with length $l$ with respect to the rest of the system, is 
$S=\frac{c}{3}\ln l+\gamma$, where $c$
is the central charge of the CFT and $\gamma$ is a cut-off dependent constant, which diverges in the absence
of an ultraviolet cutoff\cite{Holzhy,CC}. In the presence of the {\it{same cut-off schemes}}, it makes sense to say that a fixed point with larger central charge
has more bipartite entanglement than the one with the smaller central charge. The celebrated $c-$theorem in $1+1$ dimensions seems to put the 
discussion on a more solid ground\cite{Zamol,CH}. The first version of the c-theorem \cite{Zamol} is about the behavior of particular correlation functions under renormalization group (RG), 
however, the second version \cite{CH} is explicitly based on the actual behavior of the entanglement entropy under RG.
For related discussions in higher dimensions, see Refs.\cite{Cardy1988,Komargodski2011,Myers2010,Jafferis2011,CH2012}. Other information theory quantities, such as, relative entropy and Fisher information
have been also used to study the renormalization group flows in different quantum field theories, see for example Refs.\cite{Balasubramanian,CTT2017,MMS2015,Lashkari2017}.
Such kind of studies is useful in the classification of different quantum (conformal) field theories \cite{Balasubramanian}.
Apart from these studies, there are also many other, more {\it{traditional}}, studies regarding renormalization group, see Ref.\cite{Osborn2017} and references therein.
One of the results in this direction is the $\eta$-conjecture, which states that in the $\Phi^4$ theories the stable fixed point
corresponds to the fastest decay of the correlation\cite{Vicari}, see for earlier related works Refs.\cite{Michel,Brezin}. In the next sections, we will elaborate more on this conjecture
and its possible connection to the entanglement content of a field theory.

Although entanglement entropy is a perfect measure to study bipartite entanglement, there is no widely accepted multipartite entanglement measure for many body systems, 
for a review see Ref.\cite{Guhne-Toth2009}. In CFT, the only related concepts that have been studied recently are the
entanglement negativity\cite{Negativity1}, and localizable entanglement\cite{Rajabpour2015}. Recently, quantum Fisher information (QFI) is introduced, see Refs.\cite{Hyllus2012,Toth2012}, as a
quantity that certain types of multipartite entanglement can be traced from
its scaling with the system size. The Fisher information
has been known for a long time as a quantity to quantify phase parameter estimation, for a review see Ref.\cite{Pezze2014}. 
In condensed matter physics, a related quantity called fidelity has been used for more than a decade to study the quantum phase transition in
different systems\cite{Zanardi2006,CamposVenuti2007,Zanardi2007a,Zanardi2007b,You2007,Zanardi2008,Gu2008,Invernizzi2008,Gong2008,Garnerone2009, Gu,Greschner2013,Salvatori2014,Bina2016,Rams2017}.
Fidelity susceptibility has been also studied extensively in the high energy physics in the holographic context in Refs.\cite{Takayanagi2015,Lashkari,Banerjee2017,Alishahiha2017,Flory2017,Momeni2017}. Although the Fisher information (Fidelity) has been studied for many years in different areas, just recent developments
have shown that one can determine the presence of certain types of multipartite entanglement by studying
the optimization of this quantity\cite{Hyllus2012,Toth2012}.  With this application in mind, QFI has been revisited in the context of quantum phase transition, and many of
its universal features have been investigated in Ref.\cite{Hauke}. In the same work, using the  connection between QFI and dynamical susceptibility an experimental
setup is proposed to measure  this quantity, see also Ref.\cite{Macri}. Around the quantum phase transition point, QFI is universal and allows us to identify strongly
entangled phase transitions with a divergent multipartite entanglement. Naturally, one expects to also have infinite multipartite entanglement in the CFTs that describe the universality class of the 
quantum phase transition. However, similar to what we described for bipartite entanglement entropy, here, the way that QFI
diverges is related to the universal structure of the QFT. In particular, we show that the scaling operator with the smallest scaling dimension plays the most
important role. This observation and the presence of the $\eta$-conjecture makes us believe that the QFI might have interesting behavior under RG.
After establishing this connection, one can make a lot of consistency checks and also  interesting predictions. In this article, first, we review the
QFI and its connection to the multipartite entanglement entropy. Then by using the results of Ref.\cite{Hauke}, we highlight the very important role of the smallest scaling dimension in the spectrum of
the QFT. Then based on this observation, we show how one can derive some of the old conclusions and also predict new results regarding the multipartite entanglement content of
quantum (conformal) field theories.

\section{Quantum Fisher information}

Quantum Fisher information quantifies the distinguishability of the density matrix $\rho$ from the unitarily shifted probe state
$\rho(\theta)=e^{-i\theta \mathcal{O}}\rho e^{i\theta \mathcal{O}}$, for  Hermitian operator $\mathcal{O}$. The interesting result
is the quantum Cram\'er-Rao bound which states that for $M$ measurements the variance of the parameter $\theta$ is bounded by 
the QFI, i. e. $(\Delta\theta)^2\geq \frac{1}{M F_Q}$. In other words, for every outcome of the measurement on probe state, we have an estimator for $\theta$. The variance of the estimator
is bounded by the QFI, i.e. $F_Q$. For pure states, $F_Q$ has a very simple form\cite{Braunstein1994}:
\begin{eqnarray}\label{QFI pure state}
F_Q=4\Delta(\mathcal{O})^2=4(\langle\psi|\mathcal{O}\mathcal{O}|\psi\rangle-\langle\psi|\mathcal{O}|\psi\rangle)^2.
\end{eqnarray}
For mixed states, another compact formula is available, see Ref.\cite{Braunstein1994}. To connect the QFI to the multipartite entanglement content of a system, 
we first need to define k-producible pure states. Consider a state of $N$ particles, then the state $|\Psi_{k-\text{prod}}\rangle$
is k-producible if $|\Psi_{k-\text{prod}}\rangle=\otimes_{l=1}^P|\phi_l\rangle$, where $|\phi_l\rangle$'s are non-producible states of $N_l\leq k$ particles, such that $\sum_{l=1}^PN_l=N$.
A state is a genuine k-partite entangled pure state if it is k-producible but not (k-1)-producible. This definition can  also be extended to mixed states\cite{Horodecki2009}.

In the Refs.\cite{Hyllus2012,Toth2012}, it was shown that for a system of N particles with spin $\frac{1}{2}$, the QFI can detect certain types of multipartite entanglement. 
The precise statement is as follows: 
consider the operator $\mathcal{O}_{lin}=\frac{1}{2}\sum_{l=1}^N\bold{n}_l. \pmb{\sigma}_l$, where $\pmb{\sigma}_l$ is the vector of Pauli matrices and $\bold{n}_l$ is a vector
on the Bloch sphere. Optimize the QFI over all the possible choices of $\mathcal{O}_{lin}$. The remarkable result of Refs.\cite{Hyllus2012,Toth2012} is that the system has useful $k+1$-partite entanglement
if
\begin{eqnarray}\label{multipartite E}
F_Q[\rho_{k-\text{prod}}]> \lfloor\frac{N}{k}\rfloor k^2+(N-\lfloor\frac{N}{k}\rfloor k)^2,
\end{eqnarray}
where $\lfloor x\rfloor$ is the floor function. When $k$ is a divisor  of $N$ the above equation has a simple form with respect to the density of QFI 
\begin{eqnarray}\label{multipartite E density}
f_Q:=\frac{F_Q[\rho_{k-\text{prod}}]}{N}>k.
\end{eqnarray}
 Similar conclusion is also valid for systems with higher spins as far as $\mathcal{O}$
represents a sum of local operators with a bounded spectrum\cite{Pezze2014}. Application of the above theorem to quantum phase transition point has
remarkable consequences. Consider a local scaling operator $\mathcal{O}_i^\alpha$ at site $i$ with scaling dimension $\Delta_{\alpha}$, then define the global operator
$\mathcal{O}^\alpha=\sum_{i=1}^N\mathcal{O}_i^\alpha$ as we defined in the above, then one can show that \cite{Hauke}:
\begin{eqnarray}\label{QFI scaling}
f_Q^{\alpha} \asymp N^{d-2\Delta_{\alpha}}.
\end{eqnarray}
One can now optimize the density of QFI by considering the smallest scaling dimension present in the system. This will, of course, leads us to
the conclusion that if there is any relevant operator in the spectrum of the system, $f_Q $ will diverge with the system size and consequently, one can conclude that
the system has divergent multipartite entanglement. This is not surprising at all because we already explained that in QFT the entanglement measures are often divergent but the interesting
observation is that the operator with the smallest scaling dimension is the one which dictates the way that the measure diverges. In other words, in the presence of the same renormalization group scheme
, one can argue that a system with smaller {\it{smallest scaling dimension}} has {\it{more}} multipartite entanglement content. Although the above argument seems very natural, one should be careful that the proof in Refs.\cite{Hyllus2012,Toth2012} considers a discrete system with particular conditions. One should be careful that elevating the validity
of such arguments to the quantum field theories is not a trivial thing. We will come back to this point again in the next section.

\section{$\eta$-conjecture and possible generalizations}
In the previous section, we highlighted the important role of the smallest scaling dimension present in the system. In the field theories written in the Ginzburg-Landau 
form, it seems natural to expect that the operator with the smallest scaling dimension is the Ginzburg-Landau field $\Phi$ itself. Based on the $\eta$-conjecture  
we have, \cite{Vicari}:

{\tn {$\eta$-conjecture:}{\it{ In general $\Phi^4$ theories with a single quadratic invariant, the infrared stable FP is the one that corresponds to the fastest decay of correlations.}}

Based on this conjecture the scaling dimension of the $\Phi$ field in the $\Phi^4$ theories goes uphill. If we assume that this field is the operator with the smallest scaling dimension 
and all the argument in Refs.\cite{Hyllus2012,Toth2012} can be generalized to $\Phi^4$ theories, then one can argue that the multipartite entanglement entropy decreases
under the RG flow. It is tempting to try to generalize the above conjecture to more generic cases. One possible generalization in two dimensions
is as follows:

{\tn {Conjecture: } {\it{The smallest scaling dimension in the spectrum of a  system always increases under renormalization group between
two unitary diagonal conformal fixed points.}}

We support this fact using Polyakov's one loop conformal perturbation theory, see for example Refs.\cite{Cardy,Delfino}. Consider a conformal fixed point
perturbed by an operator $\phi$ (and corresponding coupling $g_{\phi}$) with scaling dimension $\Delta_{\phi}$, which is the least relevant scaling operator
in the spectrum of the system.
Since the operator
with the smallest scaling dimension $\mathcal{O}$ does not mix with the other operators the $\beta$-functions can be written as:
\begin{eqnarray}\label{beta functions}
\beta(g_{_{\phi}})&=&(d-\Delta_{\phi})g_{_{\phi}}-c_{_{\phi\phi\phi}}g_{_{\phi}}^2-c_{_{\mathcal{O}\mathcal{O}\phi}}g_{_{\mathcal{O}}}^2+...,\\
\beta(g_{_{\mathcal{O}}})&=&(d-\Delta_{\mathcal{O}})g_{_{\mathcal{O}}}-c_{_{\mathcal{O}\mathcal{O}\mathcal{O}}}g_{_{\mathcal{O}}}^2-c_{_{\mathcal{O}\mathcal{O}\phi}}g_{_{\mathcal{O}}}g_{_{\phi}}+...,
\end{eqnarray}
where $c_{ijk}$'s are the structure constants of the CFT. Because of the perturbation the RG flow takes the system to a new fixed point with 
$(g_{_{\phi}}^*,g_{_{\mathcal{O}}}^*)=(\frac{d-\Delta_{\phi}}{c_{_{\phi\phi\phi}}},0)$. In the new fixed point, the conformal weight of the smallest scaling dimension is
\begin{eqnarray}\label{smallest scaling dimension}
\Delta'_{\mathcal{O}}=\Delta_{\mathcal{O}}+(d-\Delta_{\phi})\frac{c_{_{\mathcal{O}\mathcal{O}\phi}}}{c_{_{\phi\phi\phi}}}+....
\end{eqnarray}
The second term is positive if the structure constants are both positive or negative. In diagonal $1+1$ dimensional CFT's, it is already proven that the structure constants
 are all positive for unitary CFTs, see Ref.\cite{DF1985}. 
 Then based on the equation (\ref{smallest scaling dimension}), one can conclude that up to one loop calculations, the smallest scaling dimension goes uphill under RG. 
 Note that in all of our discussion, we just consider massless perturbations, which take the system from a non-trivial CFT to another non-trivial CFT. These are
 the cases that the Polyakov conformal perturbation theory can be applied safely.
 The other important fact is that one can not use the above argument for generic scaling operators simply because they usually mix with the other operators
under the RG and, so they have very different forms at different fixed points. Note that our analysis is reminiscent of the famous $\Delta$-theorem discussed in Ref.\cite{Delfino}.
 
 Having the above result, one can now argue that the quantum Fisher information and consequently, the multipartite entanglement entropy decreases under RG flow very similar to what 
 we have for bipartite
 entanglement entropy\cite{CH}.
\subsection { 1+1d diagonal CFTs: Ginzburg-Landau description }
A field theory which fits perfectly to our line of argument is the Ginzburg-Landau  description of unitary minimal models\cite{DMS1997} with the Lagrangian
\begin{eqnarray}\label{LG lagrangian}
\mathcal{L}=\int d^2z\{\frac{1}{2}(\partial\Phi)^2+\Phi^{2(m-1)}\},
\end{eqnarray}
with the central charge $c(m)=1-\frac{6}{m(m+1)}$. The operator with the smallest scaling dimension is $\Phi$, which 
corresponds to  the operator $\phi_{2,2}$ in the Kac table with the conformal dimension $\Delta_{2,2}=\frac{3}{2m(m+1)}$. It is not difficult to see that
$\Delta_{22}=\frac{1-c}{4}$. This simple analysis shows that one expects bigger entanglement content for systems with larger central charges, because they have smaller smallest scaling dimension.
Also it is quite well-known that perturbing the Ginzburg-Landau Lagrangian with the relevant operator $\Phi^{2(m-2)}$ takes the system from the fixed point with the central charge $c(m)$
to the fixed point with the central charge $c(m-1)$ which is smaller, however, since $\Delta_{22}(m-1)>\Delta_{22}(m)$, we expect less entanglement at the end of the RG flow. This picture is
perfectly consistent with the famous result regarding bipartite entanglement entropy which follows the formula $S=\frac{c}{3}\ln l+\gamma$, see Ref.\cite{CH}. One should notice that our line of argument is
radically different from the common arguments because here, instead of emphasizing on the behavior of the central charge, we are giving more importance to the smallest
scaling dimension in the system. 
\subsection { 1+1 dimensional non-diagonal CFTs:}
Note that the above results are true for any QFT that can be described by the $\mathcal{A}$-series of the minimal unitary CFTs. In general, it is not true 
that any CFT with the bigger central charge has smaller {\it{smallest scaling dimension}}. For example, a CFT in $\mathcal{D}$-series with bigger central 
charge might have a bigger smallest scaling dimension
than a CFT in the $\mathcal{A}$-series. In addition, two CFTs in different series might have the same central charge but different smallest scaling dimensions\cite{DMS1997}. The most famous one
is the conformal field theory with the central charge $c=\frac{4}{5}$ which describes $Q=3$-states Potts model. We will discuss this model in more detail later.
It is quite interesting to see what prevents us to extend the conjecture of the last section to non-diagonal cases. First of all, the structure constants 
in the $\mathcal{D}\mathcal{E}$-series are not always non-negative\cite{Petkova,Fuchs,Zuber}, however, it seems that the structure constants appearing in the OPE of the smallest scaling dimension with
the rest of the operators can be chosen positive. This, however, does not guaranty that the smallest scaling dimension always goes up under RG. For example,
in the $\mathcal{D}$-series there are two copies of one operator which although their structure constants in particular basis
can be chosen non-negative, they can have negative structure constants in other relevant basis. For particular perturbation of these CFTs, the latter basis is the one which should be considered
in the calculations. The most famous example is possibly a conjectured flow from $\mathcal{D}_4$ (non-diagonal $Q=3$-states Potts model) to $A_4$ (tri-critical Ising model)
discussed in Refs.\cite{Ravanini,Klassen1993}. This counter example forces us to take a closer look to the concept of the operator content in more detail.
\section { Operator content in QFT, CFT and discrete model:}
Having a discrete model, it is normally very difficult to find the full operator content of the system, especially if the system is not integrable.
The problem 
is more tractable in two dimensional CFTs. In two dimensions, when one talks about operator content of a CFT it means
that in the torus partition function of the model, the characters of certain operators are appearing. For example, in the Ising CFT partition function on the torus the operators $\epsilon$
and $\sigma$ play the important role. Now consider the partition function of the discrete Ising model on the torus. This partition function is proportional to the Ising CFT partition function
that we
just discussed.
Although,  in CFTs on the torus, the operator content has a well-defined definition, it is not necessarily a {\it{full description}} of the discrete model.
A discrete model with different boundary conditions can lead to different CFTs on the torus. On top of that, it is possible to define different operators for the discrete model
and study their correlations, but the characters of these operators do not necessarily appear in the torus partition function. The same is true also when one studies a Lagrangian
QFT. In the next subsection, we will discuss a concrete example. In Ref.\cite{Balasubramanian}, for earlier similar discussion see  Ref.\cite{Douglas2013}, one can find a related interesting discussion regarding
the proximity of quantum field theories and their operator content. In Ref.\cite{Balasubramanian}, the authors define a theory which is called {\it{master UV theory}}, which can be a discrete model or a continuum CFT
in a way that its deformation leads us to various low energy effective field theories. The idea is based on labeling the operator content based on the master theory. This concept seems to be useful
for our discussion regarding the quantum Fisher information and the entanglement content. The idea is based on this fact that one can always starts with a master theory and finds the operator
with the smallest scaling dimension. The character of this operator might not appear in the partition function, but it can be defined and used to detect the entanglement for the discrete model.

\subsection {$Q=3$-states Potts model:}
The quantum $Q=3$-states Potts model is a very interesting model to discuss some aspects of the arguments regarding the operator content
of a QFT and a discrete model. We first define the Hamiltonian of the discrete critical quantum model as
\begin{eqnarray}\label{Q=3 discrete Hamiltonian}
H=-J\sum_j(\sigma_{j+1}^{\dagger}\sigma_j+\sigma_{j}^{\dagger}\sigma_{j+1})-J\sum_j (\tau_j^{\dagger}+\tau_j),
\end{eqnarray}
where the operators on different sites commute but on the same sites, we have $\sigma_j^3=\tau_j^3=1$ and $\sigma_j\tau_j=\omega\tau_j\sigma_j$
with $\omega=e^{2\pi/3}$. As it is clear the Hamiltonian has $Z_3$ symmetry. 
The {\it{operator content}} of this model for different boundary conditions was discussed in Ref.\cite{Cardy1986}. Instead of going through all 
the possibilities, we stick to just cases that are useful for our discussion. For periodic boundary conditions(PBC), the operators that appear in the partition function are the ones with
dimensions $(0,0)$ (identity operator $\mathcal{I}$), $(\frac{2}{5},\frac{2}{5})$ (energy operator $\epsilon$), $(\frac{7}{5},\frac{7}{5})$ (operator $X$),
$(3,3)$ (operator Y), $(3,0)$ and $(0,3)$ (operators $\Phi_{3,0}$ and $\Phi_{0,3}$), $(\frac{7}{5},\frac{2}{5})$ and $(\frac{2}{5},\frac{7}{5})$ (operators $\Phi_{\frac{7}{5},\frac{2}{5}}$ and
 $\Phi_{\frac{2}{5},\frac{7}{5}}$) and two copies of the operators with dimensions $(\frac{1}{15},\frac{1}{15})$ (operators $\sigma$ and $\sigma^{\dagger}$)
 and, $(\frac{2}{3},\frac{2}{3})$ (operators $Z$ and $Z^{\dagger}$). This CFT is called the $\mathcal{D}_4$, and, as it can be seen here, the smallest scaling dimension is $(\frac{1}{15},\frac{1}{15})$.
 The lattice form of these operators can be written explicitly with respect to lattice parafermions, spins and dual spins, see Ref.\cite{Fendley2014}.
 
 In the twisted boundary condition (TBC), different operators with different scaling dimensions starts to show up\cite{Cardy1986}, including
 $(\frac{1}{8},\frac{1}{8})$, $(\frac{1}{40},\frac{1}{40})$, $(\frac{21}{40},\frac{21}{40})$, $(\frac{13}{8},\frac{13}{8})$, $(\frac{13}{8},\frac{1}{8})$, $(\frac{1}{8},\frac{13}{8})$,
 $(\frac{21}{40},\frac{1}{40})$ and, $(\frac{1}{40},\frac{21}{40})$. They can be labeled as $R_{a,b}$ where $(a,b)$ is the scaling dimension of the operator.
 The operators $R_{\frac{1}{8},\frac{1}{8}}$ and $R_{\frac{1}{40},\frac{1}{40}}$ are called disorder operators, see Ref.\cite{Zamolodchikov} and their presence is attributed to
 this fact that the Hamiltonian is actually symmetric with respect to the dihedral group $D_6$ which is equivalent to $(Z_3,\tilde{Z}_3)$. This extra part comes from this fact that 
 the Hamiltonian is also invariant under charge conjugation. As it is clear the smallest scaling dimension in this sector is $(\frac{1}{40},\frac{1}{40})$, which is also the 
 case for the diagonal CFT $\mathcal{A}_6$. In the $\mathcal{A}_6$ CFT, we have all the scaling spinless operators that we introduced so far\cite{DMS1997}. The disorder operators
  $R_{\frac{1}{40},\frac{1}{40}}$ and $R_{\frac{1}{8},\frac{1}{8}}$, can be defined for the Hamiltonian (\ref{Q=3 discrete Hamiltonian}) as a 
  string of charge conjugation operators\cite{Cardy,Zamolodchikov,Fendley2014}. This lattice operators can be defined independent of the boundary conditions, and so in some sense they are
  present even if they do not appear explicitly in the partition function. For example, when one discusses the quantum Fisher information, they can be used to detect
  the multipartite entanglement. We note that apparently their non-local nature is not an obstacle \cite{Lipori2017}. The above discussion means that if we take
  periodic lattice $Q=3$-states Potts model, then, the smallest scaling dimension of an operator that we can define, has dimension $(\frac{1}{40},\frac{1}{40})$
  but the character of this operator is absent in the partition function. Similar argument seems to be valid for the Ginzburg-Landau 
  representation of the $\mathcal{D}_4$ model. In this case, the Ginzburg-Landau field theory  has the following form\cite{Cardy1986,Li}
 \begin{eqnarray}\label{Li version}
S^*=\int d^dr[(\partial\Phi_1)^2+(\partial\Phi_2)^2+\frac{\lambda}{\sqrt{2}}(\Phi_1^3-3\Phi_1\Phi_2^2)],
\end{eqnarray}
which after redefinition $\Phi=(\phi_1+i\phi_2)/\sqrt{2}$ and $\Phi^*=(\phi_1-i\phi_2)/\sqrt{2}$ can be also written as
\begin{eqnarray}\label{Cardy version}
S^*=\int d^Dr[|\partial\Phi|^2+\lambda(\Phi^3+\Phi^{*3})].
\end{eqnarray}
In this form, the $Z_3$ symmetry is more manifest. In this field theory, the two copies of the spin operator $\sigma$ are $\Phi$ and $\Phi^*$
operators with conformal dimension $h=\frac{1}{15}$. The other two copies of the spin operator $Z$ are 
\begin{eqnarray}\label{Z operators}
\Phi_{13}^+&=&\frac{\Phi^{*2}\Phi+\Phi^{2}\Phi^*}{\sqrt{2}}=\frac{\phi_1^3+\phi_1\phi_2^2}{2},\\
\Phi_{13}^-&=&\frac{-\Phi^{*2}\Phi+\Phi^{2}\Phi^*}{\sqrt{2}i}=\frac{\phi_2^3+\phi_1^2\phi_2}{2},
\end{eqnarray}
with conformal dimension, $h=\frac{2}{3}$. Obviously, in this theory, the Ginzburg-Landau field has the dimension $(\frac{1}{15},\frac{1}{15})$
which as we discussed in the previous section could be the operator with the smallest scaling dimension. However, as we discussed for the discrete case, 
one might be able to define a charge conjugation string operator which has smaller scaling dimension. Although this has not been investigated in detail,
the lessons taken from the discrete model support the idea that the smallest scaling dimension might be this string operator.
Now consider the following perturbation of the field theory in the equation(\ref{Li version}):
\begin{eqnarray}\label{perturbation}
S=S^*+g\int d^Dr \Phi_{13}^+.
\end{eqnarray}
It is conjectured in Ref.\cite{Klassen1993} that the field theory after perturbation
 flow to a new fixed point, which is in the universality class of the tri-critical Ising model with the central charge $c=\frac{7}{10}$. Note that, the above perturbation can be done 
 explicitly also for the discrete model.
In the same paper, it is argued that, the conformal dimension of $\phi_1$ and $\phi_2$ at the new fixed point are  $\frac{7}{16}$ and $\frac{3}{80}$. If this is true that means that
the scaling dimension of $\phi_2$ is actually getting smaller under RG in contradiction to what we have in the $\eta$-conjecture for the $\Phi^4$ theories.
The conclusion is that the dimension of the  Ginzburg-Landau field might not always increase under RG but the value of the smallest scaling dimension possibly always increases under RG.
Assuming that there is a {\it{UV master theory}} to start to explore different low energy effective QFTs, it is tempting to make the following statement:

{\tn { } {\it{Starting from a master UV theory the smallest scaling dimension in the spectrum of a  system always increases under renormalization group between
two unitary  conformal fixed points.}}

An equivalent statement is to say that the QFI decreases under renormalization group.
\subsection {The cut effect:}

Another important issue (often overlooked) in the study of the entanglement entropy in QFTs, is the effect of the cut \cite{Holzhy,Tachikawa2015,Rajabpour2016,Oxman2017}.
Consider a quantum spin chain, then, it is easy to say that one is interested in the entanglement of one part of the chain with respect to the rest. However,
in the continuum field theory, the boundary between two regions is not well defined. Normally, one needs to consider a small UV cutoff size region 
between the two domains that we would like to calculate their entanglement. However, then one needs to consider a particular boundary condition there. The nature of 
this boundary condition depends on the form of the cut. In the discrete models, it comes naturally, but in a field theory, it is more obscure in general. In $1+1$ dimension,
the cut forces us to work with the partition function on the annulus with particular boundary conditions, see Refs.\cite{Holzhy,Rajabpour2016}. The effect of the boundary conditions
on the entanglement is always 
subleading. For example, in $1+1$ dimensional CFTs, the entanglement entropy of a domain with size $l$ with respect to the rest is, \cite{Rajabpour2016} and references therein:
\begin{eqnarray}\label{rajab2015}
S=\frac{c}{6}\ln\frac{l(l+s_1)}{s_2s_1}+\ln b_1+\ln b_2+\frac{b_1^2}{b_0^2}(\frac{s_2s_1}{2l(l+s_1)})^{2\Delta_1}+ ...,
\end{eqnarray}
where $s_1$ and $s_2$ are the size of the regions at the boundary of the two domains and $b_{1,2}$ are related to the boundary 
conditions on the cuts and the corresponding terms are called Affleck-Ludwig boundary entropy. In the last term, $\Delta_1$ is the smallest scaling dimension that 
appears in the partition function of the annulus. Here again, we encounter the smallest scaling dimension, but in a subleading term. However, this time it is quite clear that
the operator with the smallest scaling dimension is the one which appears explicitly in the annulus partition function.

\section{Bipartite Entanglement Measures }
In this section, we discuss few more examples that show, knowing the smallest  scaling dimension in a system can lead to statements regarding the entanglement content of the discrete model or the QFT. 

\subsection{1+1 dimensional CFTs with $c=1$}
The models with the central charge $c=1$ are very interesting because they normally have a critical line with changing critical exponents.
The perfect examples are the compactified bosons on a circle or on orbifold with radius $r$. Since in these models the central charge is the same, the bipartite entanglement entropy
of a segment in the leading order is the same all along the critical line, however,  the subleading terms are  controlled by the smallest 
scaling dimension $\Delta_1=\frac{1}{2}min(r^2,\frac{1}{4r^2})$ as $S=\frac{1}{3}\ln l+c_1+b\frac{1}{l^{4\Delta_1}}$ with positive $b$, see Ref.\cite{Alba2011}. This means that 
after subtracting the leading term,
one can see that the critical points with smaller smallest scaling dimension have bigger entanglement entropy. 
This fact is numerically checked in the case of the Ashkin-Teller model in Ref.\cite{Alba2011}. 
This argument is correct also in the level of the mutual information
of two disjoint intervals. Note that, the subleading terms in every critical model are controlled by the smallest scaling dimension present in the system independent of the central charge, see for example Ref.\cite{Rajabpour2016} and references therein.

\subsection{Entanglement negativity in 1+1 dimensional CFT}
Entanglement negativity has been used recently to study the entanglement entropy in tripartite  many body systems \cite{negativity2} and CFT\cite{Negativity1}. The idea goes as follows:
Consider a tripartition  $A\cup B\cup \bar{B}$ of a system which is in the pure sate $\rho=|\psi\rangle\langle\psi|$ and then, trace out part $A$ of the 
system, i.e. $\rho_{_{B\cup \bar{B}}}=\text{tr}_A\rho$; finally calculate the (logarithmic) entanglement negativity (LEN) of $B$ with respect to $\bar{B}$ defined as:
\begin{eqnarray}\label{Negativity}
\mathcal{E}_{_{B:\bar{B}}}=\ln\text{tr}|\rho_{_{B\cup \bar{B}}}^{T_2}|,
\end{eqnarray}
where $\rho_{_{B\cup \bar{B}}}^{T_2}$ is the partially transposed reduced density matrix with respect to $\bar{B}$.
The LEN of two adjacent intervals 
with lengths $l_1$ and $l_2$ is \cite{Negativity1}: $\mathcal{E}=\frac{c}{4}\ln\frac{l_1l_2}{l_1+l_2}+\gamma_2$ which is just dependent on the central charge and, so it 
is compatible with our line of argument. The $c=1$, and two disjoint intervals are more interesting because one has a line of critical exponents. Based on our argument,
the LEN should be bigger for critical points with smaller smallest scaling dimension. This is apparently consistent with the numerical calculations
available for the R\'enyi version of the  LEN
performed on the XXZ chain in Ref.\cite{Alba2013}. We conjecture that it is true also for the logarithmic entanglement negativity itself.

\subsection{Localizable entanglement}

Localizable entanglement (LE) is another measure of multipartite entanglement first studied in Refs.\cite{localizable1,localizable2}, and it is based on localizing entanglement in two sections by performing projective measurements in other parts.
 The localizable entanglement
between the two parts $B$ and $\bar{B}$ after performing local projective measurement in the rest of the system $A$ is defined as:
\begin{eqnarray}\label{localizable}
E_{loc}(B,\bar{B})=sup_{\mathcal{E}}\sum_i p_i E(\ket{\psi_{i}}_{B\bar{B}}),
\end{eqnarray}
where $\mathcal{E}$ is the set of all possible outcomes $(p_i,E(\ket{\psi_{i}}_{B\bar{B}})$ of the measurements, and $E$ is the chosen entanglement measure.
The maximization is done with respect to all the possible observables to make the quantity independent of the observable. In Refs.\cite{Rajabpour2015}, and \cite{Rajabpour2016}, we found a lower bound for the localizable entanglement when the chosen measure is the von Neumann entropy.
When the two regions $B$ and $\bar{B}$ are adjacent we have
\begin{eqnarray}\label{localizable lower bound 1}
S_{loc}(B,\bar{B})>\frac{c}{6}\ln\frac{l(l+s)}{a s}+\gamma_2,
\end{eqnarray}
where $s$ and $l$ are the sizes of the regions $A$ and $B$. Since again the dominant term is proportional to the central charge, all of the previous discussions are valid. However,
the situation is more interesting when the two regions $B$ and $\bar{B}$ are completely decoupled and far from each other. In this case, we have\cite{Rajabpour2016}:
\begin{eqnarray}\label{localizable lower bound 2}
S_{loc}(B,\bar{B})>(\frac{l}{8s})^{2\Delta}\ln\frac{l}{8s},
\end{eqnarray}
where $\Delta$ is the smallest scaling dimension present in the spectrum of the system. The localizable entanglement of two disjoint regions are controlled
by the smallest scaling dimension present in the spectrum of the system. This is in perfect consistency with the behavior of the QFI. 
\subsection{Non-conformal scale invariant systems}
Our argument based on QFI is independent of the conformal symmetry and in principle, it should also be valid for the scale invariant but not conformal invariant systems.
In other words, our argument should also work  for systems that the Lorentz invariance is lost.
Here, we first discuss the  coupled long-range harmonic oscillators with the Hamiltonian in the momentum space:
\begin{eqnarray}\label{coupled HO}
H=\sum_k\frac{1}{2}\pi_k\pi_{-k}+\frac{1}{2}\omega^2(k)\phi_k\phi_{-k},
\end{eqnarray}
where $\omega^2(k)=|k|^{\alpha}$ with $0<\alpha\leq2 $. Scaling exponent of the operator $\phi$ which is the operator with the smallest scaling dimension is $\Delta_{\phi}=\frac{2-\alpha}{4}$.
Based on this exponent, one can argue that the entanglement content of the systems with smaller $\alpha$ should be smaller than the oscillators with bigger $\alpha$'s.
It was shown numerically in Ref.\cite{Ghaseminezhad} that in $1+1$ dimension, the entanglement entropy of a subsystem with length $l$
follows the formula: 
\begin{eqnarray}\label{coupled HO EE}
S=\frac{c(\alpha)}{3}\ln l+\gamma(\alpha),
\end{eqnarray}
where $c(\alpha)$ increases monotonically from zero up to one. This is remarkably consistent with our argument based on QFI and a very non-trivial check of what we have discussed so far.
For results regarding $\alpha>2$ see Ref.\cite{molla}.
Similar numerical calculations are also performed on the  long-range Ising chain in Ref.\cite{Vodola2016} with the Hamiltonian 
\begin{eqnarray}\label{Long-range ising}
H=\sin\theta\sum_{i=,j>i}^L\frac{\sigma_i^x\sigma_j^x}{|i-j|^{\alpha}}+\cos\theta\sum_{i=1}^L\sigma_i^z,
\end{eqnarray}
where $\theta$ and $h$ are some parameters. In paramagnetic phases (so called PM2 in Ref.\cite{Vodola2016}), the scaling exponent of $\sigma^x$  decreases with increasing $\alpha$.
Since $\sigma^x$  is the operator with the smallest scaling dimension we expect that the $c(\alpha)$ decreases with increasing $\alpha$. This is in contrast to the long-range coupled harmonic oscillators
that we discussed above. Remarkably the numerical calculations of Ref.\cite{Vodola2016} confirm this expectation perfectly. Similar arguments are also valid for long-range Kitaev chain
investigated in Ref.\cite{Vodola2016}.

\section{Conclusions}

In this paper, using quantum Fisher information (QFI), we studied the multipartite entanglement entropy
in conformal field theories and argued that systems with larger central charges have more entanglement content than the systems with the smaller central charges. 
We showed this by studying the smallest scaling dimension in the {\it{spectrum}} of the system. We also mentioned that the concept of the operator
content of a QFT can be a very delicate problem. Some the conclusions
regarding the bipartite (von Neumann) and multipartite entanglement (LEN and LE) entropies can be understood in a unified framework by studying
QFI. This quantity  is much useful when one is interested in comparing
the entanglement content of two or more different models. In particular, it can be very useful, if one thinks about it in the context of RG.
We believe most of the conclusions in this paper can be extended more or less straightforwardly to higher dimensions. One example is the mutual information of
two spheres which is controlled by the smallest scaling dimension as discussed in Ref.\cite{Cardy2013}. There are also other quantum
information measures that are cutoff independent and are related to
entanglement entropy, see \cite{Casini2009b,Casini2011b,Casini2015b}. It would be interesting to study these measures in the language of QFI.

\textbf{Acknowledgment:}
I thank G. Delfino, S. Rychkov and E. Vicari for discussions regarding the behavior of the smallest scaling dimension in QFT.
I thank V. Petkova and J Fuchs for useful correspondence regarding the sign of the structure constants in two dimensional CFTs.
I also thank M. Maghrebi, K. Najafi and J. Viti for  useful discussions and correspondence. 
The work of MAR was supported in part by CNPq.










\begin{thebibliography}{44}
\bibitem{Holzhy}  C. Holzhey, F. Larsen and F. Wilczek, Nucl. Phys. B {\bf{424}}, 443 (1994).

\bibitem{Sorkin} L. Bombelli, R. K. Koul, J. H. Lee and R. D. Sorkin, Phys. Rev. D {\bf{34}}, 373 (1986).

\bibitem{CW} C. G. Callan and F. Wilczek, Phys. Lett. B {\bf{333}}, 55 (1994).

\bibitem{CC}  P. Calabrese and J. Cardy, J. Stat. Mech. 06 (2004) {\bf{P06002}}.

\bibitem{Zamol}A. B. Zamolodchikov, Pis’ma Zh. Eksp. Teor. Fiz. {\bf{43}}, 565 (1986); JETP Lett. 43, 730 (1986)

\bibitem{CH} H. Casini and M. Huerta, Phys. Lett. B {\bf{600}}, 142 (2004).

\bibitem{Cardy1988} J. L. Cardy,  Phys. Lett. B {\bf{215}} (1988) 749–752.

\bibitem{Komargodski2011} Z. Komargodski and A. Schwimmer, J. High Energy Phys. {\bf{12}} (2011) 099.

\bibitem{Myers2010} R. C. Myers and A. Sinha,  Phys. Rev. D {\bf{82}} (2010) 046006.

\bibitem{Jafferis2011} D. L. Jafferis, I. R. Klebanov, S. S. Pufu, and B. R. Safdi, J. High Energy Phys. {\bf{06}} (2011) 102.

\bibitem{CH2012} H. Casini and M. Huerta,  Phys. Rev. D {\bf{85}} (2012) 125016.




\bibitem{Balasubramanian}     V. Balasubramanian, JJ. Heckman and A Maloney, JHEP {\bf{05}},  104 (2015).
\bibitem{CTT2017}  H. Casini, E. Testé and G. Torroba, JHEP {\bf{03}}, 089 (2017).
\bibitem{MMS2015}  R. Maity, S. Mahapatra and T. Sarkar, Phys. Rev. E {\bf{92}},  052101 (2015).

\bibitem{Lashkari2017} N. Lashkari,  [arXiv:1704.05077].

\bibitem{Osborn2017} H. Osborn, A. Stergiou, [ arXiv:1707.06165 ]
\bibitem{Vicari} E. Vicari, J. Zinn-Justin,  New J.Phys. {\bf{8}} (2006) 321 
\bibitem{Michel} L. Michel, Phys. Rev. B {\bf{29}} (1984) 2777.
\bibitem{Brezin} K. Brezin, J. C. Le Guillou and J. Zinn-Justin, Phys. Rev. B {\bf{10}}, 892(1974).

\bibitem{Guhne-Toth2009} O. G\"{u}hne and G. T\`{o}th, Phys. Rep. {\bf{474}}, 1 (2009).

\bibitem{Negativity1} P. Calabrese, J. Cardy, E. Tonni, Phys. Rev. Lett. {\bf{109}}, 130502 (2012) 
and  J. Stat. Mech. (2013) {\bf{P02008}}. 

\bibitem{Rajabpour2015} M. A. Rajabpour, Phys. Rev. B {\bf{92}}, 075108 (2015).

\bibitem{Hyllus2012}  P. Hyllus, W. Laskowski, R. Krischek, C. Schwemmer, W. Wieczorek, H. Weinfurter, L. Pezz\`{e},
and A. Smerzi, Phys. Rev. A {\bf{85}}, 022321 (2012).

\bibitem{Toth2012} G. T\`{o}th, Phys. Rev. A {\bf{85}}, 022322 (2012).

\bibitem{Pezze2014} L. Pezz\'e and A. Smerzi,  
in {\it{Atom Interferometry, Proceedings of the International School of Physics Enrico Fermi, Course 188, Varenna (eds Tino, G. and Kasevich, M.)}} 691-741 (IOS Press, 2014).

\bibitem{Zanardi2006}  P. Zanardi and N. Paunkovic, Phys. Rev. E {\bf{74}}, 031123 (2006).
\bibitem{CamposVenuti2007}  L. Campos Venuti and P. Zanardi, Phys. Rev. Lett. {\bf{99}}, 095701 (2007).
\bibitem{Zanardi2007a} P. Zanardi, M. Cozzini, P. Giorda, J. Stat. Mech. (2007) L02002.
\bibitem{Zanardi2007b} M. Cozzini, P. Giorda, P. Zanardi, Phys. Rev. B {\bf{75}}, 014439 (2007).
\bibitem{You2007}  W.-L. You,  Y.-W. Li,   and S.-J. Gu, Phys. Rev. E {\bf{76}}, 022101 (2007).
\bibitem{Zanardi2008}  P. Zanardi, M. G. A. Paris,  and L. Campos Venuti, Phys. Rev. A {\bf{78}}, 042105 (2008).
\bibitem{Gu2008}  S.-J. Gu, H.-M. Kwok, W.-Q. Ning,  and H.-Q. Lin, Phys. Rev. B {\bf{77}}, 245109 (2008).
\bibitem{Invernizzi2008}  C.  Invernizzi,  M.  Korbman,  L.  Campos  Venuti,    and M. G. A. Paris, Phys. Rev. A {\bf{78}}, 042106 (2008).
\bibitem{Gong2008}  L. Gong and P. Tong, Phys. Rev. B {\bf{78}}, 115114 (2008).

\bibitem{Garnerone2009} S. Garnerone, D. Abasto, S. Haas, P. Zanardi, Phys. Rev. A {\bf{79}}, 032302 (2009). 
\bibitem{Gu} S-J. Gu,  Int. J. Mod. Phys. B {\bf{24}}, 4371  (2010).
\bibitem{Greschner2013}  S. Greschner, A. K. Kolezhuk,  and T. Vekua, Phys. Rev. B {\bf{88}}, 195101 (2013).
\bibitem{Salvatori2014}  G. Salvatori, A. Mandarino,   and M. G. A. Paris, Phys. Rev. A {\bf{90}}, 022111 (2014).
\bibitem{Bina2016}   M. Bina, I. Amelio,   and M. G. A. Paris, Phys. Rev. E {\bf{93}}, 052118 (2016).

\bibitem{Rams2017} M.M. Rams, P. Sierant, O. Dutta, P. Horodecki, J. Zakrzewski [arXiv:1702.05660].




\bibitem{Takayanagi2015} M. Miyaji, T. Numasawa, N. Shiba, T. Takayanagi, K. Watanabe, Phys. Rev. Lett. {\bf{115}}, 261602 (2015).
\bibitem{Lashkari} N. Lashkari, M. Van Raamsdonk,  JHEP {\bf{1604}}, 153 (2016).
\bibitem{Banerjee2017} S Banerjee, J Erdmenger, D Sarkar [1701.02319].
\bibitem{Alishahiha2017} M. Alishahiha, A. Faraji Astaneh, [arXiv:1705.01834].
\bibitem{Flory2017} M. Flory, JHEP {\bf{06}}, 131  (2017).
\bibitem{Momeni2017} D. Momeni, M. Faizal, K. Myrzakulov, R. Myrzakulov, Phys. Lett. B {\bf{765}}, 154 (2017).  




\bibitem{Hauke} P. Hauke, M. Heyl, L. Tagliacozzo, and P. Zoller, Nat. Phys. {\bf{12}}, 778 (2016).

\bibitem{Macri} T. Macri,  A. Smerzi, L. Pezze, Phys. Rev. A {\bf{94}}, 010102(R) (2016)

\bibitem{Braunstein1994}  S. L. Braunstein and C. M. Caves, Phys. Rev. Lett. {\bf{72}}, 3439 (1994).

\bibitem{Horodecki2009}  R. Horodecki, P. Horodecki, M. Horodecki, and K. Horodecki, Rev. Mod. Phys. {\bf{81}}, 865 (2009).



\bibitem{Cardy} J. Cardy, {\it{Scaling and renormalization in statistical
physics}} (Cambridge University Press, Cambridge, 1996).
 
\bibitem{Delfino} G. Delfino, P. Simonetti, J.L. Cardy, Phys.Lett. B {\bf{387}}  327 (1996)
 \bibitem{DF1985} Vl. S. Dotsenko and V.A. Fateev, Physics Letters B {\bf{154}}, 291(1985).
 A.B. Zamolodchikov and V.A. Fateev, Sov. J. Nucl.Phys. {\bf{43}} 657 (1986) and
 P. Christe and R. Flume, Phys. Lett. B {\bf{188}} 21 (1987)

 \bibitem{DMS1997} P. Di Francesco, P. Mathieu, D. Senechal,
{\it{Conformal Field Theory}} , Graduate Texts in Contemporary Physics, Springer-Verlag, New York, (1997).

 \bibitem{Petkova}V. B. Petkova, Int. J. Mod. Phys. A {\bf{3}}  2945 (1988); V. B. Petkova, Phys. Lett. B {\bf{225}}, 357 (1989); P. Furlan, A. Ch. Ganchev and V.B. Petkova,
Int. J. Mod. Phys. A {\bf{5}}  2721 (1990); Erratum, ibid. 3641.

\bibitem{Fuchs}  J. Fuchs, Phys. Rev. Lett. {\bf{62}} 1705 (1989); 
J. Fuchs and A. Klemm, Ann.Phys. (N.Y.) {\bf{194}}, 303 (1989); J. Fuchs, Phys. Lett. B {\bf{222}}, 411  (1989); J. Fuchs, A. Klemm und C. Scheich, Z. Phys.C {\bf{46}},  71 (1990).

\bibitem{Zuber} V. B Petkova, J. B Zuber, Nucl. Phys. B {\bf{438}}, 347 (1995)
\bibitem{Klassen1993}  T. Klassen and E. Melzer, Nucl. Phys. {\bf{350}}, 635 (1991)
 
\bibitem{Ravanini} F. Ravanini, Phys. Lett. B {\bf{274}}  345 (1992).

\bibitem{Douglas2013} M. R. Douglas,  J. Phys. Conf. Ser. {\bf{462}}, 012011 (2013).

\bibitem{Cardy1986}  J. L. Cardy, Nucl. Phys. B {\bf{275}}, 200 (1986)
\bibitem{Zamolodchikov}  A. B. Zamolodchikov and V. A. Fateev, Sov. Phys. JETP {\bf{63}}, 913 (1986).
\bibitem{Fendley2014} R. S. K. Mong, D. J. Clarke, J. Alicea, N. H. Lindner, P. Fendley, J. Phys. A: Math. Theor. {\bf{47}} 452001 (2014) 


\bibitem{Lipori2017} L. Pezzè, M. Gabbrielli, L. Lepori, A. Smerzi, [arXiv:1706.06539 ]
\bibitem{Li}K. Li, Phys. Lett. B {\bf{219}} 297 (1989) 



\bibitem{Tachikawa2015} K. Ohmori, Y. Tachikawa,  J. Stat. Mech. (2015) P04010. 
\bibitem{Rajabpour2016} M. A. Rajabpour, J. Stat. Mech. (2016) {\bf{063109}} and K. Najafi, M. A. Rajabpour, JHEP {\bf{1612}} (2016) 124.
\bibitem{Oxman2017} D. R. Junior and L. E. Oxman, Phys. Rev. D {\bf{95}}, 125005 (2017).




\bibitem{Alba2011}V. Alba, L. Tagliacozzo, P. Calabrese,  J. Stat. Mech.(2011) 1106:{\bf{P06012}}.



\bibitem{negativity2} 
A. Bayat, P. Sodano, S. Bose, Phys. Rev. B {\bf{81}}, 064429 (2010).
A. Bayat, S. Bose, P. Sodano, H. Johannesson, Phys. Rev. Lett. {\bf{109}}, 066403 (2012).

\bibitem{Alba2013} V. Alba, J. Stat. Mech. (2013) {\bf{P05013}}.



\bibitem{localizable1}  F. Verstraete, M. Popp and J. I. Cirac,  Phys. Rev. Lett. {\bf{92}}, 027901(2004); 

\bibitem{localizable2}  F. Verstraete, M.A. Martin-Delgado, J.I. Cirac;
. Phys. Rev. Lett. {\bf{92}}, 087201 (2004) and M. Popp, F. Verstraete, M. A. Martin-Delgado, J. I. Cirac
Phys. Rev. A {\bf{71}}, 042306 (2005)

\bibitem{Ghaseminezhad} M. Ghasemi Nezhadhaghighi, M. A. Rajabpour, EPL, {\bf{100}} (2012) 60011 and Phys. Rev. B {\bf{88}}, 045426 (2013)

\bibitem{molla} M Mozaffar, A Mollabashi, [arXiv:1705.00483]

\bibitem{Vodola2016} D. Vodola, L. Lepori, E. Ercolessi and G. Pupillo, New J. Phys. 18 {\bf{015001}} (2016)

\bibitem{Cardy2013} J. Cardy,  J. Phys. A: Math. Theor. {\bf{46}}  285402 (2013)
\bibitem{Casini2009b} H. Casini and M. Huerta JHEP {\bf{03}} (2009) 048 
\bibitem{Casini2011b} D. D. Blanco and H. Casini, Class. Quantum Grav. {\bf{28}}  215015 (2011)  
\bibitem{Casini2015b} H. Casini, M. Huerta, R.C. Myers, JHEP {\bf{10}} (2015) 003






















\end{thebibliography}
\end{document}